\documentclass[]{spie}  

\usepackage{amsmath,amsfonts,amssymb}
\usepackage{graphicx}
\usepackage{color}
\usepackage[colorlinks=true, allcolors=blue]{hyperref}

\title{Optical Kerr spatiotemporal dark extreme waves}

\author[a,b]{Stefan Wabnitz}
\author[c]{Yuji Kodama}
\author[a,b]{Fabio Baronio}

\affil[a]{Dipartimento di Ingegneria dell'Informazione, Universit\`a di Brescia, via Branze 38, 25123, Brescia, Italy}
\affil[b]{Istituto Nazionale di Ottica del Consiglio Nazionale delle Ricerche (INO-CNR), Via Branze 45, 25123 Brescia, Italy}
\affil[c]{Department of Mathematics, Ohio State University, Columbus, OH 43210, USA}

\begin{document} 
\maketitle

\begin{abstract}
We study the existence and propagation of multidimensional dark non-diffractive and non-dispersive
spatiotemporal optical wave-packets in nonlinear Kerr media.
We report analytically and confirm numerically the properties of 
spatiotemporal dark lines, X solitary waves and lump solutions of the  $(2+1)D$  nonlinear 
Schr\"odinger equation (NLSE). Dark lines, X waves and lumps represent holes of light on a continuous wave
background.  These solitary waves are derived by exploiting the connection between 
the $(2+1)D$ NLSE and a well-known equation of hydrodynamics, namely the $(2+1)D$ Kadomtsev-Petviashvili (KP)
equation.  This finding opens a novel path for the excitation and control of spatiotemporal optical solitary and rogue waves,
of hydrodynamic nature.
\end{abstract}


\section{INTRODUCTION}
\label{sec:intro}  
Techniques to shape and control the propagation of electromagnetic radiation are of paramount importance in many fields of basic science and applied research, such as atomic physics, spectroscopy, communications, material processing, and medicine \cite{boyd08,weiner}. Among these, of particular interest are methods that produce localized and distortion-free wave-packets, i.e. free from spatial and temporal spreading due to diffraction or material group velocity dispersion (GVD), respectively \cite{recami}. 
There are two main strategies to achieve propagation invariant electromagnetic wave packets \cite{majus14}. The first methodology is based on the spatiotemporal synthesis of a special input wave, so that diffractive and dispersive effects compensate for each other upon linear propagation
in the material. Building blocks of these linear light bullets are Bessel beams and their linear combinations, along with Airy pulses \cite{Bessel}. 
These waveforms enable the generation spatiotemporal invariant packets such as the Airy-Bessel beams, and the so-called X-waves, 
obtained by a linear superposition of Bessel beams with different temporal frequencies.
The second approach involves the generation of solitary waves, that exploit the nonlinear (quadratic or cubic) response of the material for compensating 
diffractive and dispersive wave spreading \cite{silber90,wise99}. Although successfully exploited in $(1+1)D$  propagation models, that describe
for example temporal solitons in optical  fibers and spatial solitons in slab waveguides, in more than one dimension spatiotemporal solitons 
have so far largely eluded experimental observation, owing to their lack of stability associated with the presence of modulation instability (MI), 
collapse and filamentation. 

Here, we overview our recent contributions to the field of non-diffractive and non-dispersive wave-packets in Kerr media  \cite{baro16,baro16ol,baro17odp},
by deriving analytically and confirming numerically the existence and propagation of 
novel multidimensional $(2+1) D$ dark non-diffractive and non-dispersive spatiotemporal solitons propagating 
in i) self-focusing and normal dispersion  Kerr media, and 
in ii) self-defocusing and anomalous dispersion Kerr media.
The analytical dark solitary solutions are derived by exploiting the connection between 
the $(2+1)D$ NLSE and the $(2+1)D$ Kadomtsev-Petviashvili (KP) equation \cite{KP}, a well-known 
equation of hydrodynamics.
Our results extend and confirm the connection between nonlinear wave propagation in optics and 
hydrodynamics, that was first established in the $1990$'s \cite{turi88,peli95,kivs96}.

\section{Optical NLSE--hydrodynamic KP mapping}
\label{sec:sc}
In the presence of group-velocity dispersion and one-dimensional diffraction, the dimensionless time-dependent paraxial wave equation in cubic Kerr media reads as \cite{conti03}:
\begin{equation}\label{2DNLS}
iu_z+\frac{\alpha}{2}u_{tt}+\frac{\beta}{2}u_{yy}+\gamma |u|^2 u=0,
\end{equation}
namely the  (2+1)D, or more precisely (1+1+1)D, NLSE, where $u(t,y,z)$ stands for the complex wave envelope, and $t,y$ represent the retarded time, in the frame traveling at the natural group-velocity, and the spatial transverse coordinate, respectively, and $z$ is the longitudinal propagation coordinate. Each subscripted variable in Eq. (\ref{2DNLS}) stands for partial differentiation. $\alpha, \beta, \gamma$ are normalized real 
constants that describe the effect of dispersion, diffraction and Kerr nonlinearity, respectively. 

We refer \eqref{2DNLS} to as \emph{elliptic} NLSE if $\alpha\beta>0$, and \emph{hyperbolic} NLSE if $\alpha\beta<0$.
In the case of weak nonlinearity, weak diffraction and slow modulation, the dynamics of optical NLSE dark envelopes
 $u(t,y,z)$ may be related to the hydrodynamic KP variable $\eta(\tau,\upsilon,\varsigma)$ as follows \cite{baro16,baro16ol}:
\begin{align}\label{NLSEKP}
u(t,y,z)=\sqrt{\rho_0+\eta(\tau,\upsilon,\varsigma)} \ \ e^{i[\gamma\rho_0z +\phi(\tau,\upsilon,\varsigma)]}
\end{align}
where $\rho_0$ stands for a background continuous wave amplitude,   $\eta(\tau,\upsilon,\varsigma)$ represents a small amplitude variation, say $\eta\sim \mathcal{O}(\epsilon)$ with $0<\epsilon\ll 1$ and the order one background $\rho_0$; $\phi=-(\gamma/c_0) \int  \eta(\tau,\upsilon,\varsigma) d\tau$; 
$\eta(\tau,\upsilon,\varsigma)$ satisfies the KP equation,
\begin{equation}\label{KP}
\left(-\eta_\varsigma+\frac{3\alpha\gamma}{2c_0}\eta\eta_\tau+\frac{\alpha^2}{8c_0}\eta_{\tau\tau\tau}\right)_\tau-\frac{c_0 \beta}{2\alpha}\eta_{\upsilon\upsilon}=0,
\end{equation}
where $\tau=t-c_0z, \upsilon=y$, $\varsigma=z$ with  $c_0=\sqrt{-\gamma\alpha\rho_0}$, $\alpha \gamma<0$
(see  \cite{baro16} for further details).

Of interest in the optical context, the \textit{elliptic} anomalous dispersion and self-defocusing regime \cite{baro16}
($\alpha > 0$, $\beta>0, \gamma<0$) leads to the KP-I regime, while
the \textit{hyperbolic} normal dispersion and self-focusing regime \cite{baro16ol} ($\alpha <0$, $\gamma>0, \beta>0$),
leads to the KP-II regime.

Without loss of generality, we may set the following constraints to the coefficients 
of Eq. (\ref{2DNLS}), $|\alpha|=4 \sqrt{2}, \beta= 6\sqrt{2}, |\gamma|=2\sqrt{2}$;
moreover, we fix $\rho_0=1$ (thus $c_0=4$).
Note that, with the previous relations among its coefficients, in the case ($\alpha > 0$, $\beta>0, \gamma<0$), Eq. (\ref{KP}) reduces to the 
standard KP-I form: $(-\eta_{{\varsigma}}-6 \eta\eta_\tau+\eta_{\tau\tau\tau})_\tau-3\eta_{\upsilon\upsilon}=0$. Whereas in the case ($\alpha <0$, $\gamma>0, \beta>0$),
Eq. (\ref{KP}) reduces to the standard KP-II form $(-\eta_{{\varsigma}}-6 \eta\eta_\tau+\eta_{\tau\tau\tau})_\tau+3\eta_{\upsilon\upsilon}=0$.
%
Moreover, the imposed constraints to the coefficients 
of Eq. (\ref{2DNLS}) also fix the scaling between the dimensionless variables $z,t,y$ in Eq. (\ref{2DNLS}) and the corresponding real-world quantities $Z=Z_0z,T=T_0t,Y=Y_0y$. 
The longitudinal scaling factor turns out to be $Z_0=2\sqrt{2} L_{nl}$, where $L_{nl}=(\gamma_{phys} I_0)^{-1}$  is the usual nonlinear length associated with the intensity $I_0$ of the background and $\gamma_{phys}=k_0 n_{2I}$, $n_{2I}$ being the Kerr nonlinear index and $k_0$ the vacuum wavenumber. The ``transverse" scales read as $T_0=\sqrt{k'' L_{nl}/2}$ and $Y_0=\sqrt{L_{nl}/(3k_0 n)}$, where $k''$ and $n$ are the group-velocity dispersion and the linear refractive index, respectively.

 \section{Normal dispersion and self-focusing regime: nonlinear lines and X-waves} 
\label{sec:sup} 

\begin{figure}
\begin{center}
\includegraphics[width=8cm]{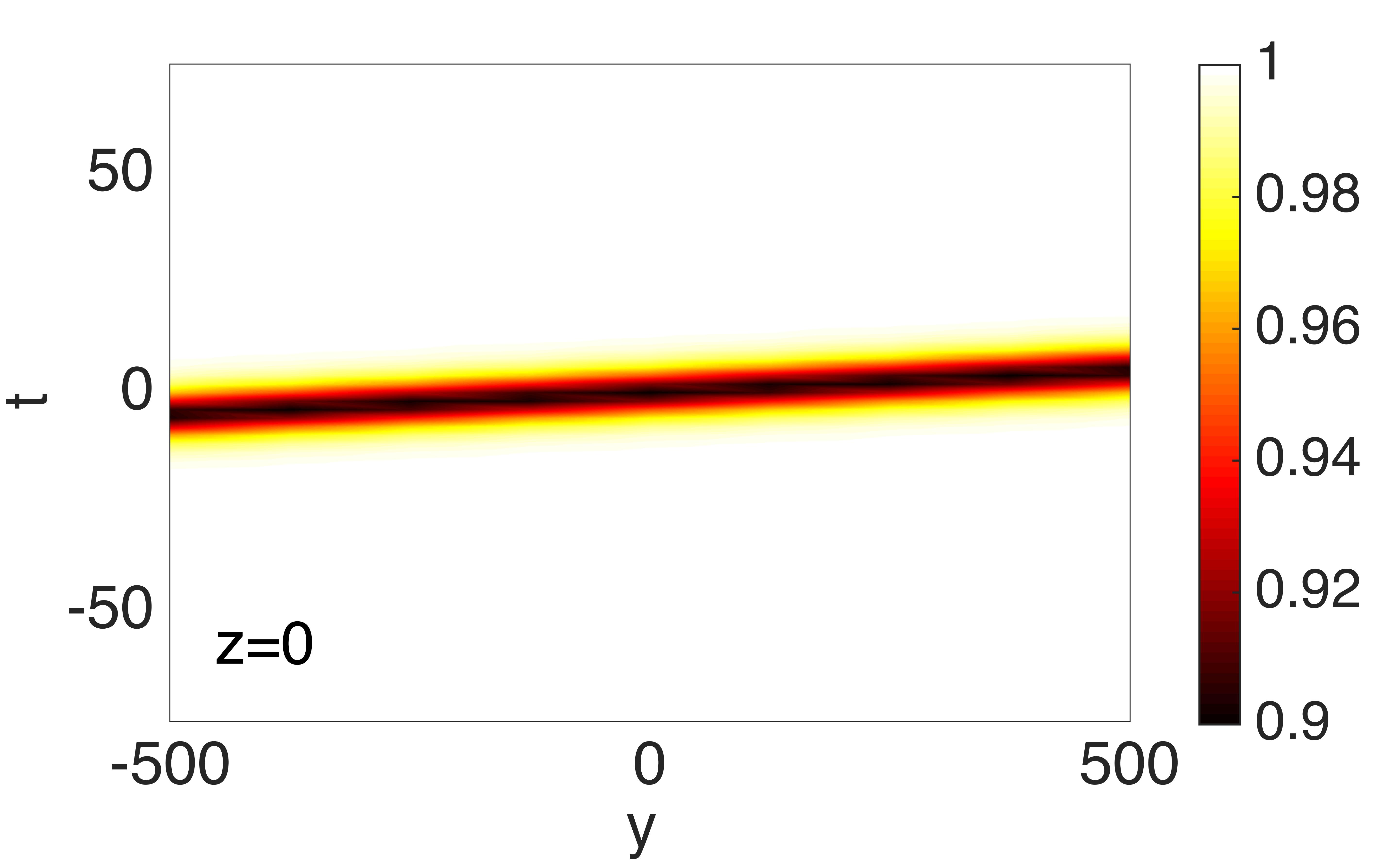}
\includegraphics[width=8cm]{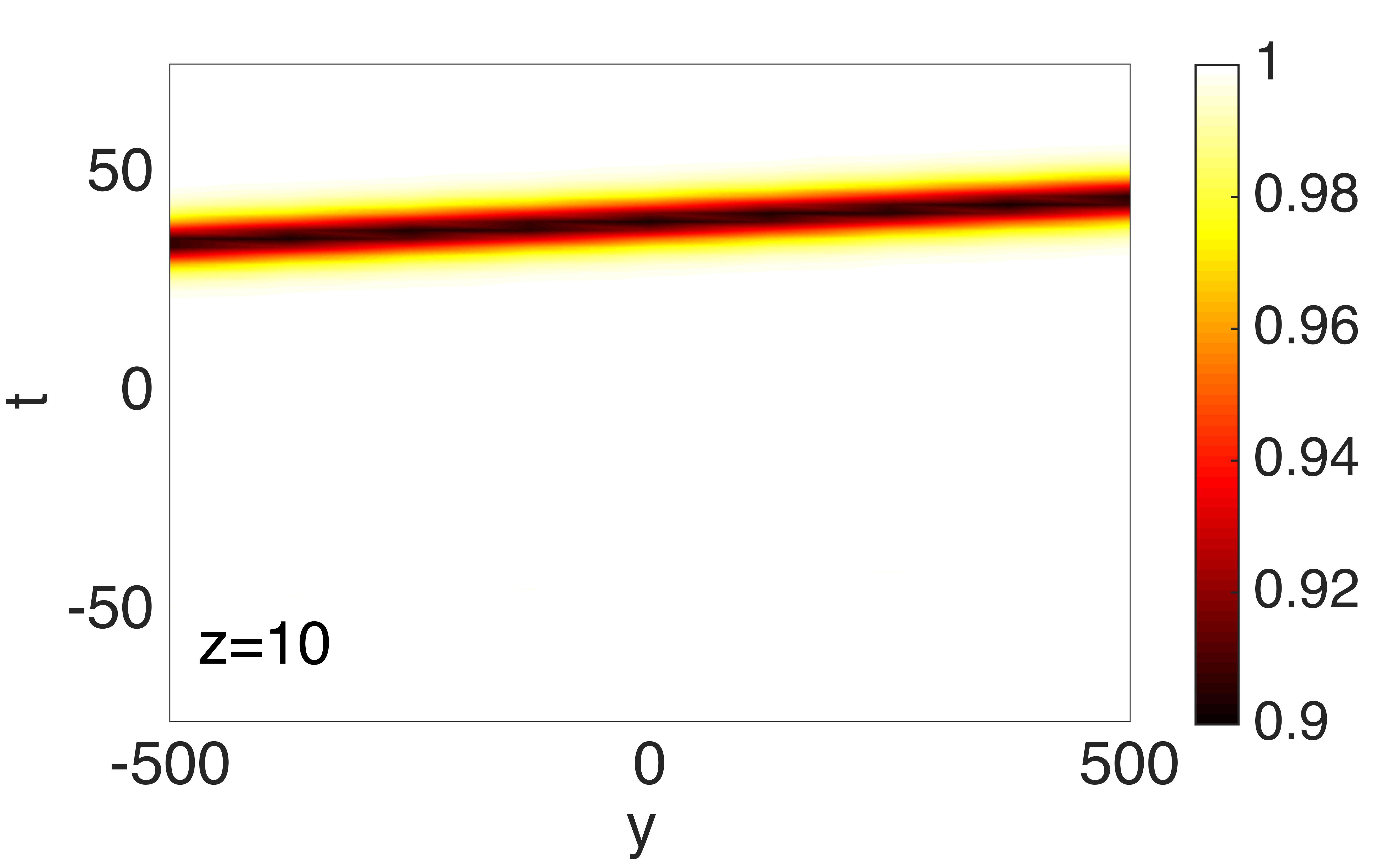}
\caption{\label{f1} Numerical spatio-temporal dark-line NLSE envelope intensity distribution
$|u(t,y,z)|^2$, shown in the $y-t$ plane, at $z=0$, at $z=10$ and in the $t-z$ plane at $y=0$. 
Here, $\epsilon=0.1$, $\varphi=0.01$. 
}
\end{center}
\end{figure}

At first, we consider the case of normal dispersion and self-focusing nonlinearity \cite{baro16ol,baro17odp}.
We proceed to consider the existence and propagation of $(2+1)D$ NLSE dark line solitary
waves, which are predicted by the existence of $(2+1)D$ KP-II bright line solitons \cite{kod10,kod11}.
When considering the small amplitude regime, a formula for an exact line bright soliton of Eq. (\ref{KP})
can be expressed as follows \cite{kod10,kod11}:
$
\eta (\tau, \upsilon, \varsigma) = - \epsilon \, \, sech^2 [\sqrt{ \epsilon/2} (\tau +tan \varphi  \, \upsilon+ c \varsigma)],
$
where $\epsilon$ rules the amplitude and width of the soliton,  $\varphi$ is the angle measured from the 
$\upsilon$ axis in the counterclockwise, $c =2 \epsilon + 3 tan^2 \varphi$ is the velocity in $\tau$-direction. 
Notice that $c$ is of order $\epsilon$. Moreover we obtain $ \phi(\tau, \upsilon, \varsigma)=\sqrt{ \epsilon} \, \, 
tanh([\sqrt{ \epsilon/2} (\tau +tan \varphi  \, \upsilon+ c \varsigma)].  $
The analytical spatiotemporal envelope intensity profile
$u(t,y,z)$ of  a NLSE dark line solitary wave is given by the mapping (\ref{NLSEKP})
exploiting the KP bright soliton expression. The intensity dip of the dark line solitary wave is $-\epsilon$, 
the velocity $c_0-c- 3tan^2 \varphi  =4-2\epsilon-3tan^2 \varphi$
in the $z$-direction. 
We numerically verified the accuracy of the analytically predicted dark line solitary waves 
of the NLSE. To this end, we made use of a standard split-step Fourier technique, which is commonly adopted 
in the numerical solution of the NLSE (\ref{2DNLS}).
Figure \ref{f1} shows the numerical spatiotemporal envelope intensity profile
$|u(t,y,z)|^2$ of  a NLSE dark line solitary wave, which corresponds to the predicted analytical dynamics.
As can be seen from the images, 
the numerical solutions of the NLSE show an excellent agreement
with the analytical approximate NLSE solitary solutions. 

In the long wave context, the KP-II equation admits complex soliton solutions, mostly discovered and
demonstrated in the last decade, 
which may describe non-trivial web patterns generated under resonances of line-solitons \cite{kod10,kod11}.
Here, we consider the resonances of four line solitons, which give birth to the so-called \textit{O-type} bright 
X-shaped two-soliton solution of the KP-II (the name \emph{O-type} is due to 
the fact that this solution was \emph{originally} found by using the Hirota bilinear method).
When considering the small amplitude regime, the formula of the  \textit{O-type} solution of 
Eq. (\ref{KP}) can be expressed as follows,
%
$\eta (\tau, \upsilon, \varsigma) = - 2 \left( \ln F\right)_{\tau\tau}, $
%
where the function $F(\tau,\upsilon,\varsigma)$ is given by $F=f_1+f_2$ with $
%
%
 %
f_1= (\epsilon_1 + \epsilon_2) \, {\rm cosh}[  (\epsilon_1 - \epsilon_2) \tau+4 \,(\epsilon_1^3 - \epsilon_2^3) \varsigma], \ \
 f_2=2 \sqrt{\epsilon_1 \epsilon_2} \,{\rm cosh}[ (\epsilon_1^2 - \epsilon_2^2)\upsilon].
$ 
$\epsilon_1, \epsilon_2$ are small real positive parameters which are related to the amplitude, width and the angle of the \textit{O-type} X-soliton solutions.
The corresponding (2+1)D NLSE dark X solitary wave $u(t,y,z)$ is directly given through the mapping Eq. (\ref{NLSEKP}),
by exploiting the soliton expression for $\eta(\tau,\upsilon,\varsigma)$.

\begin{figure}[h!]
\begin{center}
\includegraphics[width=8cm]{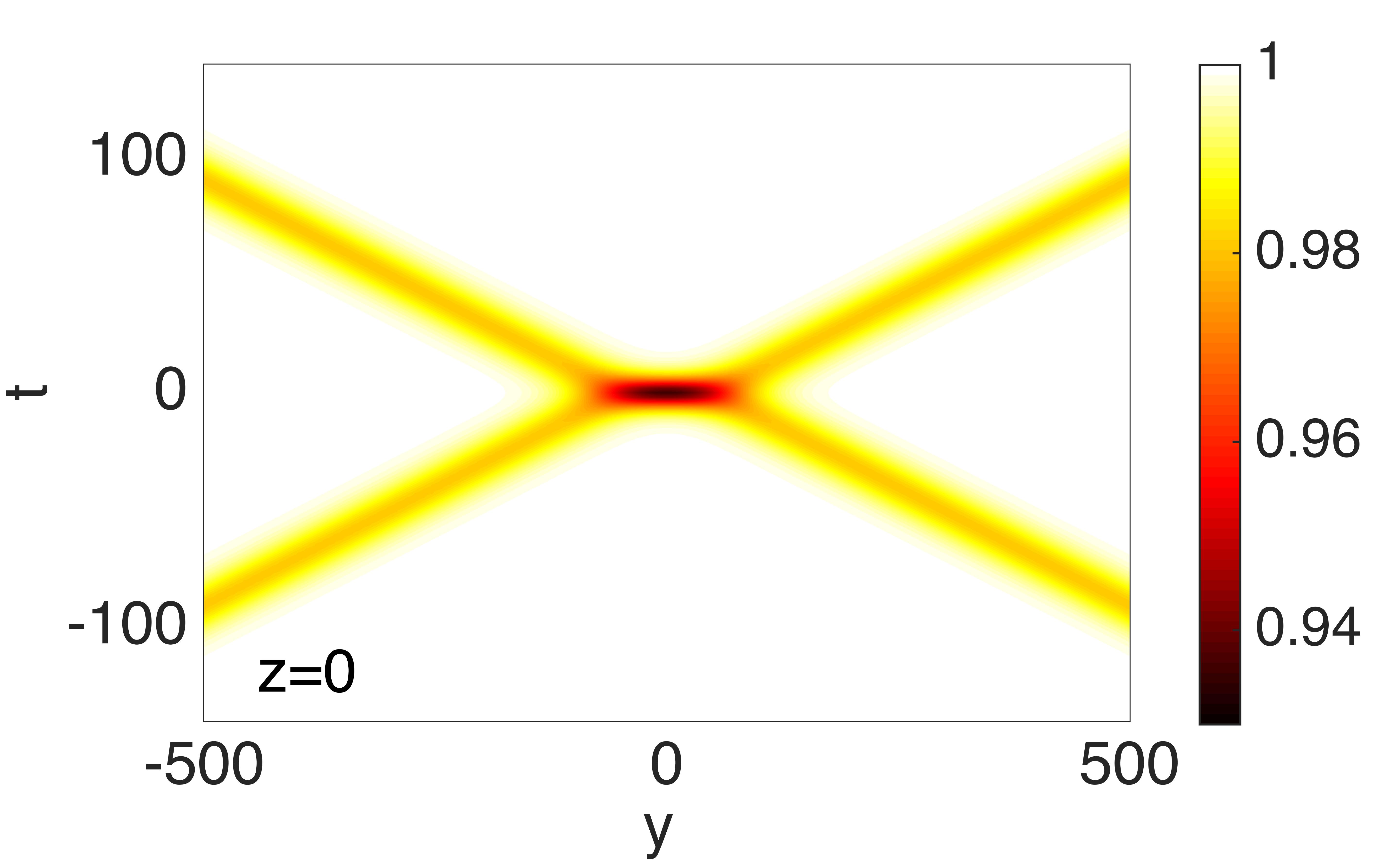}
\includegraphics[width=8cm]{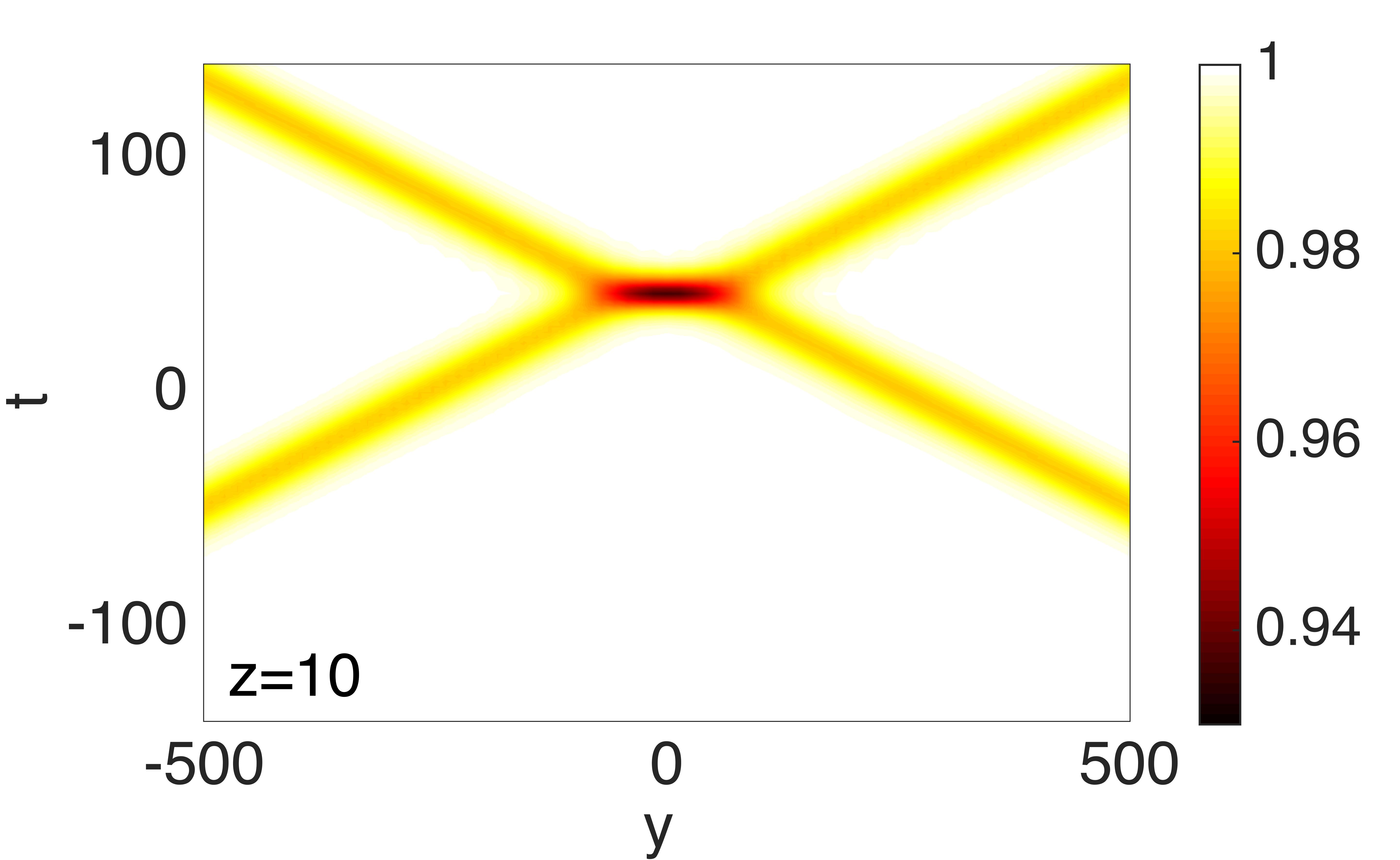}
\caption{\label{f2}  Numerical spatiotemporal NLSE envelope intensity distribution
$|u(t,y,z)|^2$, in the $(y,t)$ plane, showing the  dark X solitary wave dynamics,
  at $z=0$ and at $z=10$.
Here, $\epsilon_1=0.2$, $\epsilon_2=0.001$. 
}
\end{center}
\end{figure}

We numerically verified the accuracy of the analytically predicted \textit{O-type} dark X solitary wave of the NLSE. 
Fig. \ref{f2} shows the $(y,t)$ profile of the numerical solution of the hyperbolic NLSE at $z=0$, and at $z=10$.
In this particular example we have chosen $\epsilon_1=0.2$, $\epsilon_2=0.001$. 
Specifically, Fig. \ref{f2} illustrates a solitary solution which describes the X-interaction of four dark line solitons. 
The maximum value of the dip in the interaction region is
$2 (\epsilon_1-\epsilon_2)^2  \, (\epsilon_1+\epsilon_2) / (\epsilon_1+\epsilon_2 +2 \sqrt{\epsilon_1 \epsilon_2})$.
Asymptotically, the solution reduces to two line dark waves for $t\ll0$ and two for $t\gg0$, 
with intensity dips $\frac{1}{2}(\epsilon_1-\epsilon_2)^2$ and characteristic angles $\pm{\rm tan}^{-1}(\epsilon_1+\epsilon_2)$, 
measured  from the $y$ axis. 
Numerical simulations and analytical predictions are in excellent agreement.
We estimate the error between the asymptotic formula and the X solitary wave in the numerics to be lower than $2 \%.$

 \section{Anomalous dispersion and self-defocusing regime: dark lumps} 
\label{sec:sup} 

\begin{figure}[h!]
\begin{center}
\includegraphics[width=8cm]{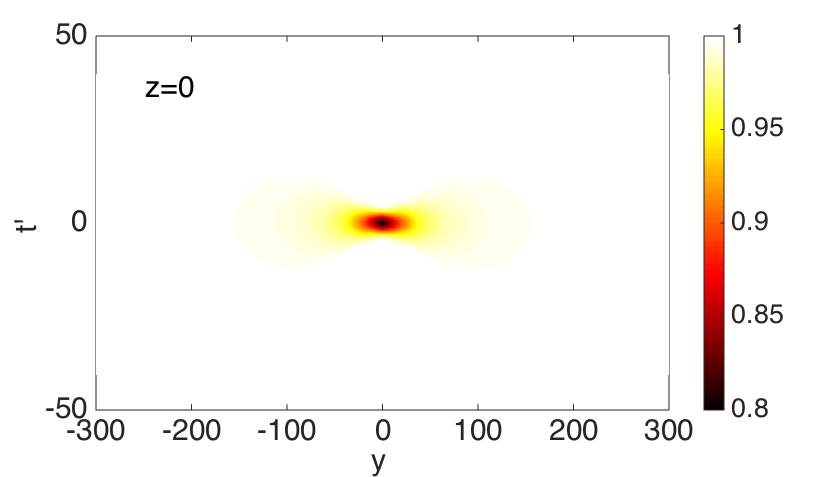}
\includegraphics[width=8cm]{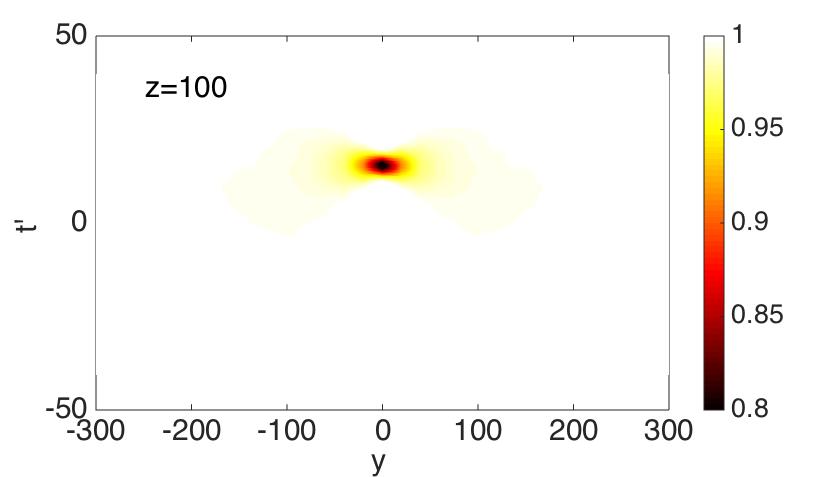}
\end{center}
\caption{\label{f3} Numerical spatio-temporal dark-lump NLSE envelope intensity distribution
$|u|^2$, shown in the $y$-$t'$ plane with $t'=t-c_0z$, at $z=0$, and $z=100$. Here, $\epsilon=0.05$. 
}
\end{figure}

Next, we consider the case of anomalous dispersion and self-defocusing nonlinearity \cite{baro16}.
We proceed to verify the existence of (2+1)D NLSE dark-lump solitary waves,
as predicted by the solutions of KP-I through Eq.(\ref{NLSEKP}) (see  \cite{as81} for details).
When considering the small amplitude regime ($\epsilon \ll 1$), a form of KP lump-soliton solution of Eq. (\ref{KP}) can be expressed as 
$\eta (\tau, \upsilon, \varsigma) = -4 [ \epsilon^ {-1}-(\tau-3 \epsilon \varsigma)^2+\epsilon \upsilon^2   ] /  [ \epsilon^ {-1}+(\tau-3 \epsilon \varsigma)^2+\epsilon\upsilon^2   ]^2$.
The parameter $\epsilon$ rules the amplitude/width and velocity properties of the KP lump soliton.
The lump peak amplitude in the $(\varsigma,\upsilon)$ plane is $-4 \epsilon$; the velocity in the  $\tau$-direction
 is $3\epsilon$. Moreover,  $\phi(\tau, \upsilon, \varsigma)= 2 \sqrt{2} \epsilon (\tau-3 \epsilon \varsigma)/ [1+ \epsilon (\tau-3 \epsilon \varsigma)+\epsilon^2 \upsilon^2].$
 The analytical spatiotemporal envelope intensity profile
$u(t,y,z)$ of  a NLSE dark solitary wave is given by the mapping (\ref{NLSEKP}), which
exploits the KP bright lump expression. Then, 
we numerically verified the accuracy of the analytically predicted dark lumps solitary waves 
of the NLSE. 
Figure \ref{f3} shows the numerical spatio-temporal envelope intensity profile
$|u|^2$ of  a NLSE dark lump solitary wave in the $y$-$t'$ plane ($t'=t-c_0 z$), 
 at the input $z=0$ and after the propagation distance $z=100$, for $\epsilon=0.05$. 
In the numerics, the initial dark NLSE profile, of KP-I lump origin, propagates stably in the $z$-direction,
with virtually negligible emission of dispersive waves,
 with the predicted velocity $c_0+3\epsilon$,
and intensity dip of $4 \epsilon$. 
Thus, the predicted theoretical dark lump solitary waves of Eq. (\ref{NLSEKP}) 
are well confirmed by numerical simulations.

\begin{figure}[h!]
\begin{center}
\includegraphics[width=8cm]{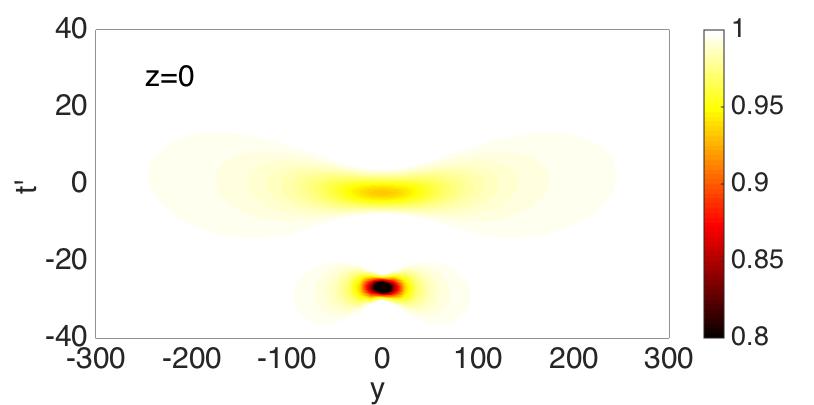}
\includegraphics[width=8cm]{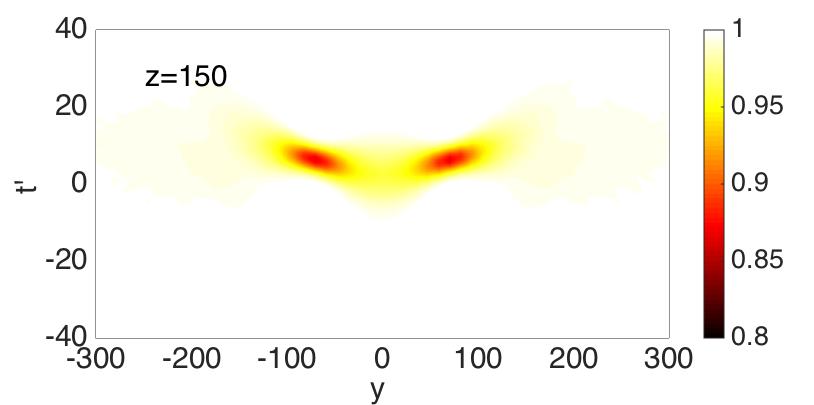}
\caption{\label{f5} Numerical spatio-temporal NLSE envelope intensity distribution
$|u|^2$, in the $y$-$t'$ plane, showing anomalous scattering of dark waves, at $z=0$, 
at $z=150$ . 
Here, $\epsilon=0.1$, $\tau_0=0, \upsilon_0=0, \varsigma_0=-50, \delta_1=0, \delta_2=0$ 
}
\end{center}
\end{figure}

We remark that the KP-I equation admits other types of lump solutions which have several peaks
with the same amplitude in the asymptotic stages $|z|\gg 0$.
We call such lump solution \emph{multi-pole lump}.
Here we show that $(2+1)$D NLSE can also support such lump solutions.
We consider multi-pole lump solution with two peaks, which is expressed as: $\eta (\tau, \upsilon, \varsigma) = -2 \partial_\tau ^2 {\rm log} F$,
where $F=|f_1^2+i f_2 +f_1/ \epsilon+1 /2 \epsilon^ {2}|^2+ |f_1+1/ \epsilon|^2 / 2\epsilon^2+1/4\epsilon^ {4}$,
and $f_1=\tau_1+2i \epsilon \upsilon_1-12 \epsilon^2 \varsigma_1 +\delta_1$, $f_2=-2 \upsilon-24i \epsilon \varsigma+\delta_2$.
$\tau_1=\tau-\tau_{0}, \upsilon_1=\upsilon-\upsilon_{0},   \varsigma_1=\varsigma-\varsigma_0$
define the dislocation; $\delta_1, \delta_2$ are arbitrary complex parameters.
The analytical spatiotemporal envelope intensity profile
$u(t,y,z)$ of  a NLSE dark solitary wave is again given by the mapping (\ref{NLSEKP}).

Figure \ref{f5} shows the initial spatio-temporal envelope intensity profile
$|u|^2$ of a two-peaked NLSE dark lump in the $y-t'$ plane, 
along with the numerically computed profiles after propagation distances $z=150$, 
for $\epsilon=0.1$ ($\tau_0=0, \upsilon_0=0, \varsigma_0=-50, \delta_1=0, \delta_2=0$). 
In particular, Fig. \ref{f5} depicts the scattering interaction of the two-peaked waves:
two dark lumps approach each other along the $t'$-axis, interact, and subsequently
recede along the $y$-axis.
These solutions exhibit anomalous (nonzero deflection angles) scattering due to multi-pole 
structure in the wave function of the inverse scattering problem.
We remark that the numerical result of NLSE dynamics 
is in excellent agreement with analytical dark solitary solution Eq. (\ref{NLSEKP})
with KP-I multi-pole lump solution.

 \section{Instabilities and Experimental Feasibility} 

Let us discuss the important issue of the stability of the predicted dark line, X solitary waves and lumps.  
Two instability factors may affect the propagation of these waves. The first one is the modulation instability (MI) of the continuous wave background.
In the case of normal dispersion and self-focusing, $\alpha <0$, $\beta, \gamma>0$, MI is of the conical type \cite{yuen80}. 
Generally speaking, MI can be advantageous to form X waves from arbitrary initial conditions both in the absence or in the presence of the background. 
However, for sufficiently long propagation distances the MI of the CW background may compete and ultimately destroy the propagation of dark solitary waves and their interactions. 
In the case of anomalous dispersion and in the self-defocusing regime,
$\alpha > 0$, $\beta>0, \gamma<0$, MI
is absent, thus lumps are not affected by MI. 

The second mechanism is related to the transverse instability of the line solitons that compose the asymptotic state of the X wave.
We point out that such instability is known to occur for the NLSE, despite the fact that line solitons are transversally stable in the framework of the KP-II (unlike those of the KP-I)  \cite{KP}. However, in our simulations of the NLSE, these transverse instabilities never appear, 
since they are extremely long-range, especially for shallow solitons. 
In fact, we found that the primary mechanism that affects the stability of dark line and X solitary waves is the MI of the CW background.

Let us briefly discuss a possible experimental setting in nonlinear optics
for the observation of cubic spatiotemporal solitary wave dynamics of hydrodynamic origin. 
As to (2+1)D spatiotemporal dynamics,  one may consider optical propagation in a planar glass 
waveguide (e.g., see the experimental set-up   of Ref. \cite{eise01}), or  a quadratic lithium
niobate crystal, in the regime of large phase-mismatch, which mimics an effective 
Kerr nonlinear regime (e.g., see the experimental set-up of Ref. \cite{baro06}). 
As far as the (2+1)D spatial dynamics is concerned, one may consider using a CW Ti:sapphire laser pulse propagating in a nonlinear medium composed of atomic-rubidium vapor (e.g., see the experimental set-up   of Ref. \cite{kivs96}),
or a bulk quadratic lithium niobate crystal, again in the regime of large phase-mismatch (e.g., see the experimental set-up of Ref. \cite{baro04, krupa15}).

 \section{Conclusions} 
We have analytically predicted a new class of dark solitary wave solutions, 
that describe non-diffractive and non-dispersive spatiotemporal localized 
wave packets propagating in optical Kerr media.
We numerically confirmed the existence of  nonlinear lines, X-waves, lumps and peculiar
 scattering interactions of the solitary waves of the (2+1)D NLSE. 
The key novel property of these solutions is that their existence 
and interactions are inherited from the hydrodynamic soliton 
solutions of the well known KP equation.
Our findings open a new avenue for research in spatiotemporal extreme nonlinear optics. 
Given that deterministic rogue and shock wave solutions, so far, have been essentially restricted to 
(1+1)D models, future research on
multidimensional spatiotemporal nonlinear waves will lead to a 
substantial qualitative enrichment of the landscape of extreme wave phenomena.

\acknowledgments 
 
We acknowledge the financial support of the European Research Council (ERC) under the European Union's Horizon 2020 research and innovation programme (grant agreement No.~740355). 


\end{document}